\pdfoutput=1
\documentclass[aps,floatfix,nofootinbib,onecolumn,superscriptaddress]{revtex4}
\usepackage{amsmath}
\usepackage{amssymb}
\usepackage{amsthm}
\usepackage{dcolumn}
\usepackage{epsfig}
\usepackage{graphics}
\usepackage{graphicx}
\usepackage{amsmath}
\usepackage{longtable}
\usepackage{color}

\definecolor{darkgreen}{rgb}{0,0.5,0}
\definecolor{purple}{rgb}{0.5,0,0.5}
\definecolor{nblue}{rgb}{0.0,0.0,0.50}
\definecolor{scarlet}{rgb}{1.0,0.2,0}
\usepackage[colorlinks=true, pdfstartview=FitV, linkcolor=purple, citecolor= purple, urlcolor=blue]{hyperref}

\begin{document}

\title{New Perspectives for QCD Physics at the LHC }

\author{STANLEY J. BRODSKY$^*$ }

\address{SLAC National Accelerator Laboratory\\
Stanford University, Stanford, California 94309 \\
and  CP$^3$-Origins, Southern Denmark University \\
Odense, Denmark\\
$^*$E-mail: sjbth@slac.stanford.edu\\}


\begin{abstract}

I review a number of topics where conventional wisdom relevant to hadron physics at the LHC  has been challenged.
For example, the initial-state and final-state interactions of the quarks and gluons entering perturbative QCD hard-scattering subprocesses lead to the breakdown of traditional concepts of factorization and universality for transverse-momentum-dependent observables at leading twist.    These soft-gluon rescattering effects produce Bjorken-scaling single-spin asymmetries, the breakdown of
the Lam-Tung leading-twist relation in Drell-Yan reactions, as well as diffractive deep inelastic scattering, 
The antishadowing of  nuclear structure functions is predicted to depend on the flavor quantum numbers of each quark and antiquark, thus explaining the anomalous nuclear dependence observed in deep-inelastic neutrino scattering.
Isolated hadrons can be produced at large transverse momentum directly within a hard higher-twist QCD subprocess, rather than from jet fragmentation, even at the LHC.   Such ``direct"  processes  can explain the observed deviations from pQCD predictions of the power-law fall-off  of inclusive hadron cross sections at  fixed $x_T= 2p_T/\sqrt s$,  as well as the ``baryon anomaly", the  anomalously large proton-to-pion ratio seen in high-centrality heavy-ion collisions at RHIC.  The intrinsic charm contribution to the proton structure function  at high $x$  can explain the large rate for  high $p_T$ photon plus charm-jet events observed by D0 at the Tevatron. The intrinsic  charm and bottom distributions imply a  large  production rate for charm and bottom jets at high $p_T$ at the LHC,  as well as a novel mechanism for Higgs and $Z^0$ production at high $x_F.$  The
light-front wavefunctions derived in AdS/QCD can be used to calculate jet hadronization at the amplitude level.
The elimination of  the renormalization scale ambiguity for the QCD coupling using
the scheme-independent BLM method will improve the precision of QCD predictions and thus increase the sensitivity of searches for new physics at the LHC. The implications of  ``in-hadron condensates"  for the QCD contribution to the cosmological constant are also discussed.

\end{abstract}

\maketitle

\vspace{10pt}

\section{Introduction}

The LHC will provide a crucial testing ground for testing QCD, not only at unprecedented  energies and momentum transfers,  but also at extreme particle densities.   In this contribution I will review a number of  topics  where unexpected new perspectives for QCD physics at the LHC have emerged.

\begin{enumerate}

\item{\it  High Transverse Momentum Hadron Production via Direct Hard Subprocesses}\\
 It is natural to assume that high transverse momentum hadrons in inclusive high energy hadronic collisions,  such as $ p p \to H X$,  can only arise  from jet fragmentation. In fact, a significant fraction of high $p^H_\perp$ isolated hadrons can emerge
directly from hard higher-twist subprocess~\cite{Arleo:2009ch,Arleo:2010yg} even at the LHC.  The direct production of hadrons can explain~\cite{Brodsky:2008qp} the remarkable ``baryon anomaly" observed at RHIC:  the ratio of baryons to mesons at high $p^H_\perp$,  as well as the power-law fall-off $1/ p_\perp^n$ at fixed $x_\perp = 2 p_\perp/\sqrt s, $ both  increase with centrality~\cite{Adler:2003kg}, opposite to the usual expectation that protons should suffer more energy loss in the nuclear medium than mesons.

A  fundamental test of leading-twist QCD predictions in high transverse momentum hadronic reactions is the measurement of the power-law
fall-off of the inclusive cross section~\cite{Sivers:1975dg}
${E d \sigma/d^3p}(A B \to C X) ={ F(\theta_{cm}, x_T)/ p_T^{n_{eff}} } $ at fixed $x_T = 2 p_T/\sqrt s$
and fixed $\theta_{CM},$ where $n_{eff} \sim 4 + \delta$. Here $\delta  =  {\cal O}(1)$ is the correction to the conformal prediction arising
from the QCD running coupling and the DGLAP evolution of the input parton distribution and fragmentation functions~\cite{Brodsky:2005fz,Arleo:2009ch,Arleo:2010yg}.
The usual expectation is that leading-twist subprocesses will dominate measurements of high $p_T$ hadron production at RHIC and Tevatron energies. In fact, the  data for isolated photon production $ p p \to \gamma_{\rm direct} X,$ as well as jet production, agrees well with the  leading-twist scaling prediction $n_{eff}  \simeq 4.5$ as shown in Fig.\ref{figNew1B} ~\cite{Arleo:2009ch}.
However, as seen in Fig.\ref{figNew1B},  measurements  of  $n_{eff} $ for hadron production  are not consistent with the leading twist predictions.
Striking
deviations from the leading-twist predictions were also observed at lower energy at the ISR and  Fermilab fixed-target experiments~\cite{Sivers:1975dg,Cronin:1973fd,Antreasyan:1978cw}.
The high values $n_{eff}$ with $x_T$ seen in the data  indicate the presence of an array of higher-twist processes, including subprocesses where the hadron enters directly, rather than through jet fragmentation~\cite{Blankenbecler:1975ct}.  The predicted deviations for the experimental and NLO scaling
exponent at RHIC and the LHC with PHENIX preliminary measurements are shown in Fig. \ref {figNew2B}.

\begin{figure}[!]
 \begin{center}
\includegraphics[width=18.0cm]{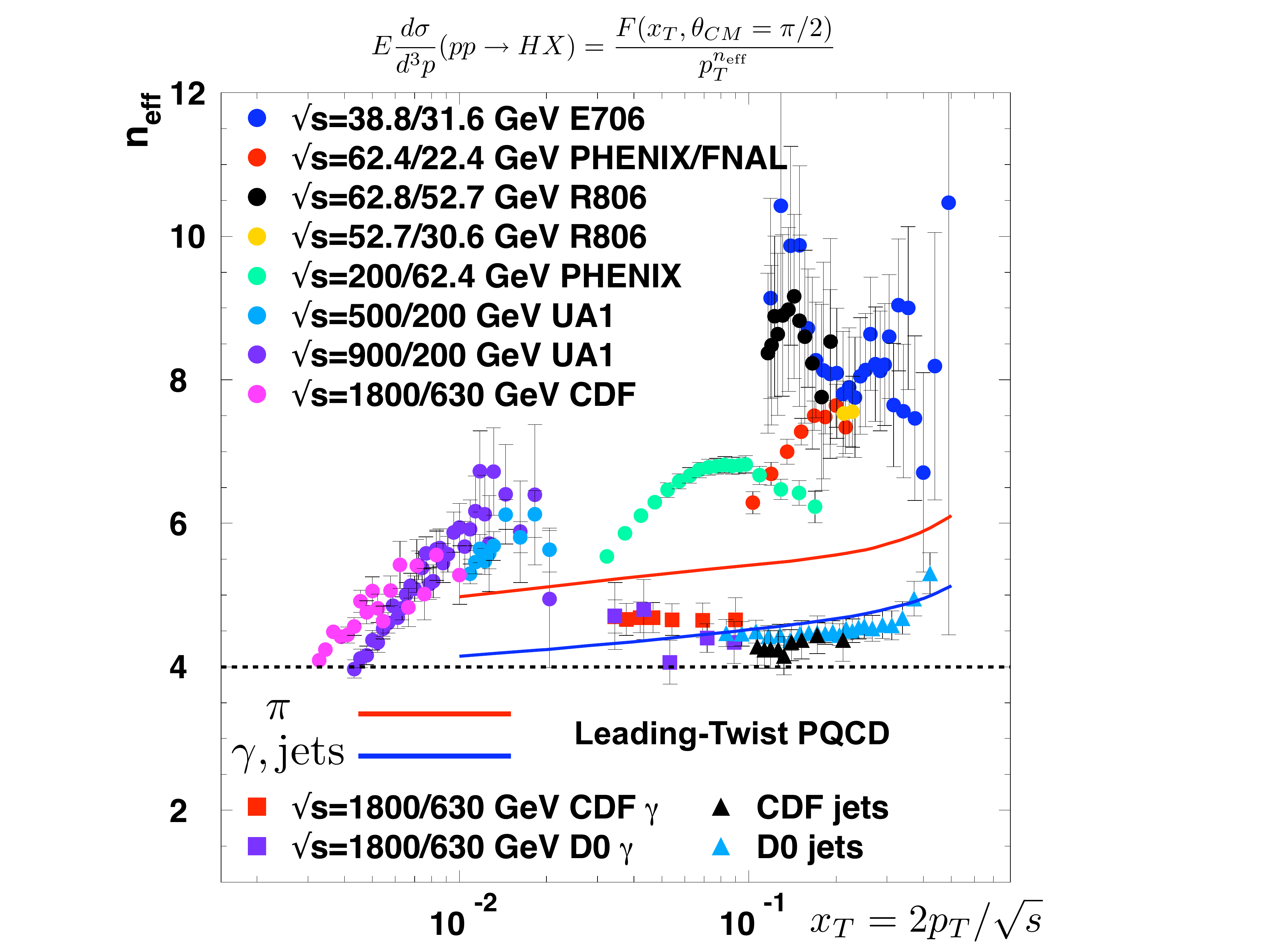}
\end{center}
  \caption{Comparison of RHIC and fixed-target data for hadron, isolated photon, and jet production with the leading-twist pQCD predictions for the power-law falloff of  the semi-inclusive cross section
  $E {d \sigma/ d^3p}(p p \to H X) = {F(x_T, \theta_{CM}=\pi/2)/p_T^{n_{\rm eff}}}$  at fixed $x_T$.
The data from R806, PHENIX, ISR/FNAL, E706 are for
charged or neutral pion production, whereas the CDF, UA1 data at small $x_T$  are for charged
hadrons. The blue curve is the prediction of  leading-twist QCD for isolated photon and jet production, including the scale-breaking effects of the  running coupling and the evolution of the proton structure functions. The red curve is the QCD prediction for pion production, which also  includes the effect from the evolution of the fragmentation function. The dashed line at $n_{\rm eff} = 4$ is the prediction of the scale-invariant parton model.   From Arleo, et al.~\cite{Arleo:2009ch}. }
\label{figNew1B}
\end{figure}

\begin{figure}[!]
 \begin{center}
\includegraphics[width=18.0cm]{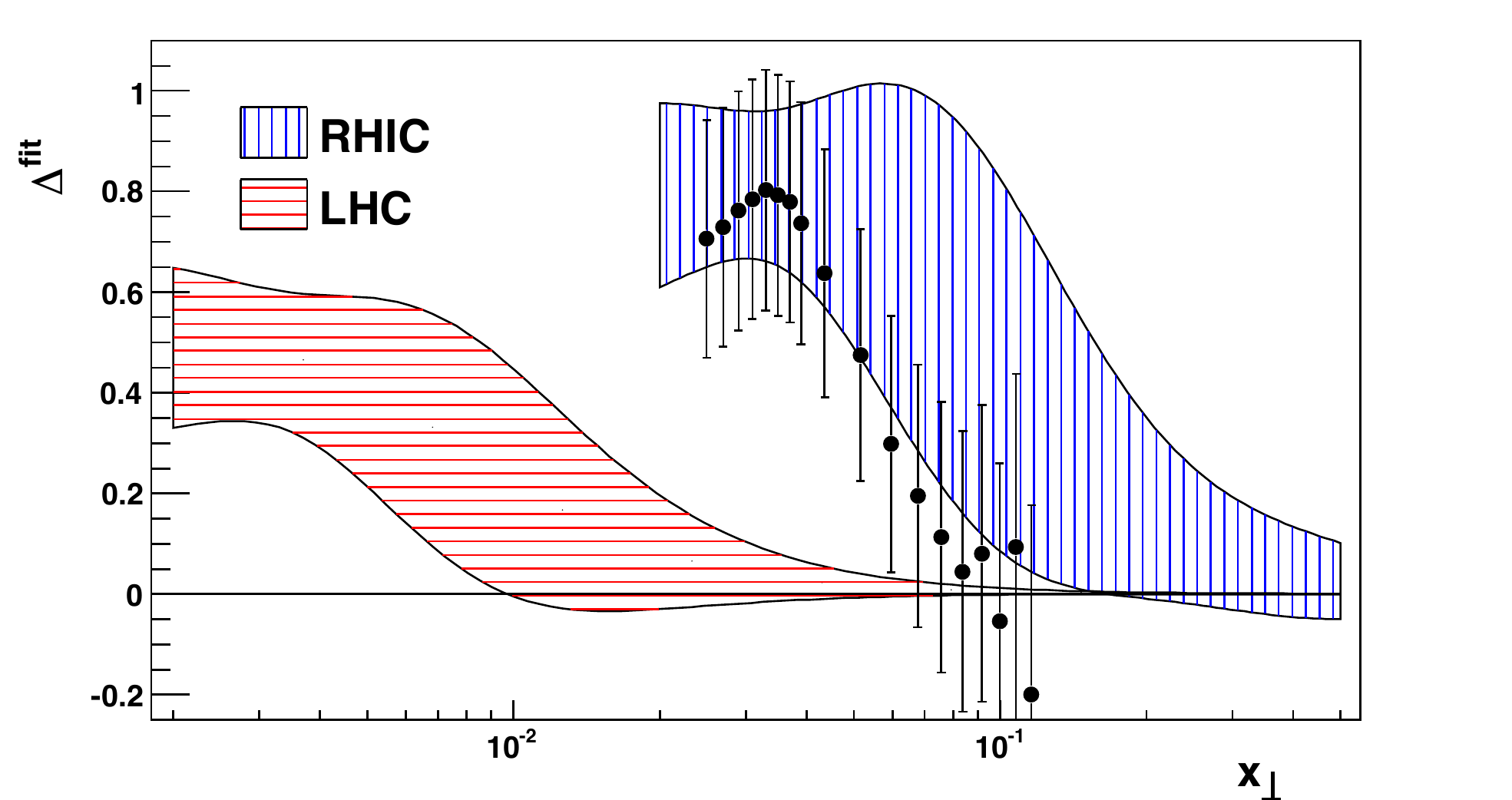}
\end{center}
  \caption{ Predicted difference $\Delta$ between the experimental and NLO scaling
exponent at RHIC ($\sqrt{s}=200,500 $ GeV) and the LHC  ($\sqrt{s}=7$~TeV as compared to $\sqrt{s}=1.8$~TeV), compared to PHENIX preliminary measurements. From Arleo, et al.~\cite{Arleo:2009ch}. }
\label{figNew2B}
\end{figure}

I will discuss further consequences of direct QCD production processes in section II.

\item{\it Breakdown of Perturbative QCD Factorization and Universality}\\
The effects of initial and final-state interactions of the quarks and gluons entering  hard processes at the LHC   are usually assumed to be of higher-twist origin and thus power-law suppressed.  However, as emphasized by
Collins and Qiu~\cite{Collins:2007nk}, the traditional factorization formalism of perturbative QCD  fails in detail for transverse-momentum-dependent observables in hard inclusive reactions because of initial- and final-state gluonic interactions at leading twist.

It is now well-understood that the final-state gluonic interactions of the scattered quark in deep inelastic lepton scattering lead to a  $T$-odd non-zero spin correlation of the plane of the lepton-quark scattering plane with the polarization of the target proton~\cite{Brodsky:2002cx}.  This  leading-twist Bjorken-scaling ``Sivers effect"  is process-dependent since QCD predicts an opposite-sign correlation~\cite{Collins:2002kn,Brodsky:2002rv} in Drell-Yan reactions due to the initial-state interactions of the annihilating antiquark.
The same final-state interactions of the struck quark with the spectators~\cite{Brodsky:2002ue}  also lead to diffractive events in deep inelastic scattering (DDIS) at leading twist,  such as $\ell p \to \ell^\prime p^\prime X ,$ where the proton remains intact and isolated in rapidity;    in fact, approximately 10\% of the deep inelastic lepton-proton scattering events observed at HERA are
diffractive~\cite{Adloff:1997sc,Breitweg:1998gc}. The presence of a rapidity gap
between the target and the diffracted proton requires that the target
remnant emerges in a color-singlet state; this is made possible in
any gauge by the soft rescattering incorporated in the Wilson line or by augmented light-front wavefunctions~\cite{Brodsky:2010vs}.

In the case of hadron-hadron collisions, the
quark and antiquark in the Drell-Yan subprocess
$q \bar q \to  \mu^+ \mu^-$ will interact with the spectators of the
other  hadron;  this leads to an anomalous $\cos 2\phi \sin^2 \theta$ planar correlation in unpolarized Drell-Yan
reactions~\cite{Boer:2002ju}.   This ``double Boer-Mulders effect" can account for the large $\cos 2 \phi$ correlation and the corresponding violation~\cite{Boer:1999mm,Boer:2002ju} of the Lam Tung relation for Drell-Yan processes observed by the NA10 collaboration.
Another  important signal for factorization breakdown at the LHC  will be the observation of a $\cos 2 \phi$ planar correlation in dijet production.

 It is usually assumed --  following the intuition of the parton model -- that the structure functions  measured in deep inelastic scattering can be computed in the Bjorken-scaling leading-twist limit from the absolute square of the light-front wavefunctions, summed over all Fock states.  In fact,  the dynamical effects, such as the Sivers spin correlation and diffractive deep inelastic lepton scattering due to final-state gluon interactions,  contribute to the experimentally observed DIS cross sections.
Diffractive events also lead to the interference of two-step and one-step processes in nuclei which in turn, via the Gribov-Glauber theory, lead to the shadowing and the antishadowing of the deep inelastic nuclear structure functions~\cite{Brodsky:2004qa};  such phenomena are not included in the light-front wavefunctions of the nuclear eigenstate.
This leads to an important  distinction between ``dynamical"  vs. ``static"  (wavefunction-specific) structure functions~\cite{Brodsky:2009dv}.

\begin{figure}[!]
 \begin{center}
\includegraphics[width=18.0cm]{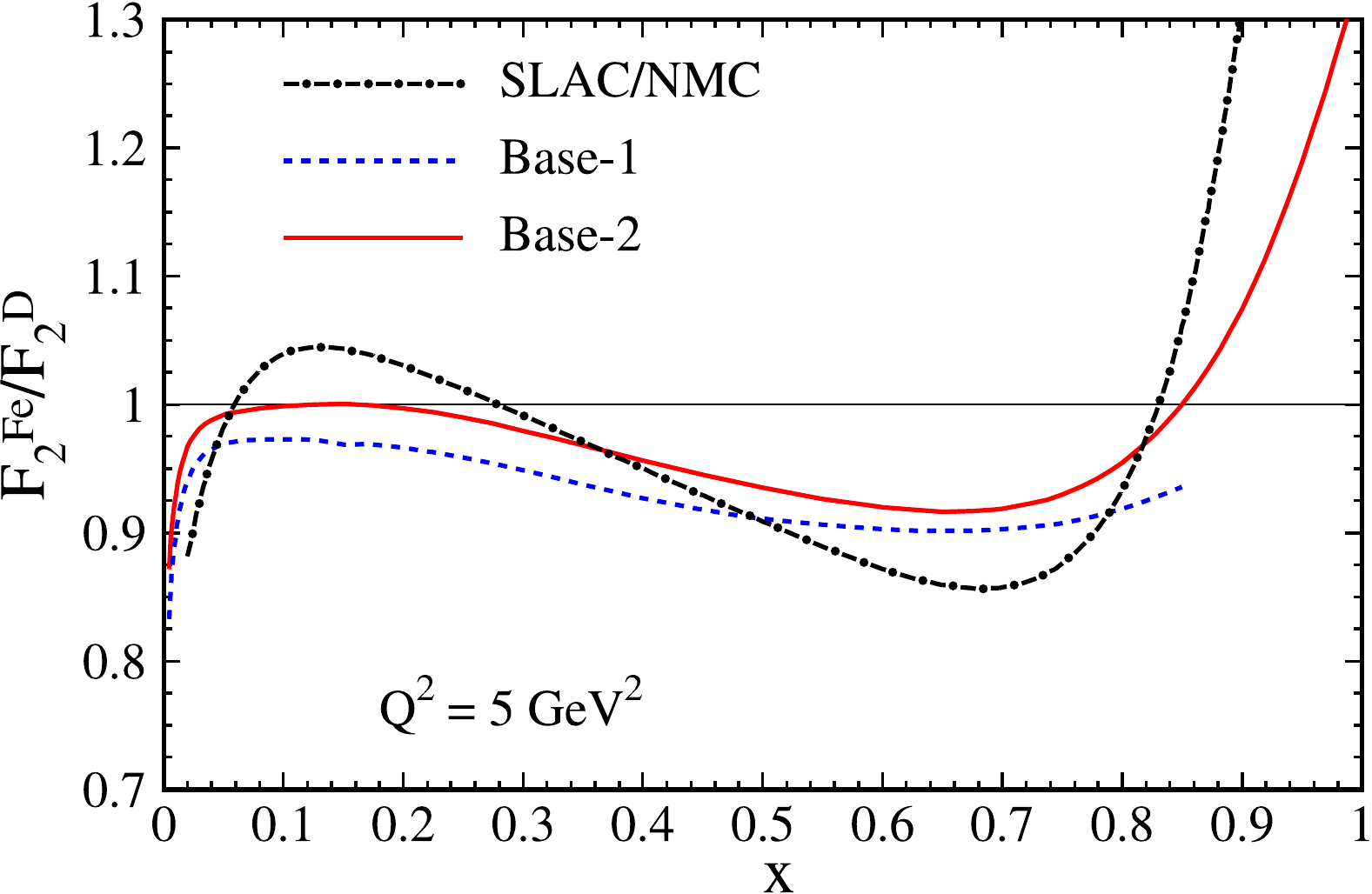}
\end{center}
  \caption{Comparison of the (Iron to Deuteron) nuclear modification  of  SLAC/NMC neutral vs.  NuTeV charged current (with range labeled by Base-1 and Base-2) deep inelastic structure functions at $Q^2=5$ GeV$^2$.  From  I.~Schienbein et al. ~\cite{Schienbein:2008ay} }
\label{figNew9}
\end{figure}

\item{\it Non-Universal Nuclear Distributions}\\
It is usually assumed  that the nuclear modifications to the structure functions measured in deep inelastic lepton-nucleus and neutrino-nucleus interactions are identical;  in fact,  the Gribov-Glauber theory predicts that the antishadowing of nuclear structure functions is not  universal, but depends on the quantum numbers of each struck quark and antiquark~\cite{Brodsky:2004qa}.   This observation can explain the recent analysis of Schienbein et al.~\cite{Schienbein:2008ay} which shows that the NuTeV measurements of nuclear structure functions obtained from neutrino  charged current reactions differ significantly from the distributions measured in deep inelastic electron and muon scattering in the $0.1 < x < 0.2$ domain.
See Fig. \ref{figNew9}.   I will discuss this in further detail in section IV.

\item{\it Intrinsic Heavy-Sea Quark Distributions}\\
Most parametrizations of the charm and bottom quark distributions in the proton structure functions only have support at low $x$ since it is conventionally assumed that they only arise from gluon splitting $g \to Q \bar Q.$  This erroneous assumption has led to many incorrect predictions; it is especially misleading for heavy hadron production at the LHC.

In fact, one can show from first principles that the proton light-front wavefunction contains {\it ab initio } intrinsic heavy quark Fock state components such as $|uud c \bar c>$~\cite{Brodsky:1980pb,Brodsky:1984nx,Harris:1995jx,Franz:2000ee}.   In contrast to the usual ``extrinsic" contribution from gluon-splitting ( i.e, DGLAP evolution), the intrinsic contributions are connected by gluons to at least two of the valence quarks of the proton. The intrinsic heavy quarks carry most of the proton's light-cone momentum since this minimizes the off-shellness of the Fock state.   One can also associate the $|uud c \bar c>$ Fock state with  meson-baryon fluctuations such as $|D(\bar c u) \Lambda_c(cud)>$.  Thus, as is the case for intrinsic strangeness, the charm and anti-charm quarks can have different momentum and spin distributions~\cite{Brodsky:1996hc}.

The probability of the intrinsic component falls as $1/M^2_{Q \bar Q}$ due to the non-Abelian QCD couplings of the gluons~\cite{Brodsky:1984nx,Franz:2000ee}.   The heavy-quark pair $Q \bar Q$ in the intrinsic Fock state  is thus primarily a color-octet,  and the ratio of intrinsic charm to intrinsic bottom scales scales as $m_c^2/m_b^2 \simeq 1/10,$ as can be verified from the operator product expansion in non-Abelian QCD~\cite{Brodsky:1984nx,Franz:2000ee}.

Intrinsic charm and bottom explain the origin of high $x_F$ open-charm and open-bottom hadron production, as well as the single and double $J/\psi$ hadroproduction cross sections observed at high $x_F$.   The factorization-breaking nuclear $A^\alpha(x_F)$ dependence  of hadronic $J/\psi$ production cross sections is also explained.   Kopeliovich, Schmidt, Soffer, Goldhaber, and I ~\cite{Brodsky:2006wb} have  proposed a novel mechanism utilizing intrinsic heavy quarks for both diffractive
 $pp \to p H p $ and inclusive Higgs production $pp \to  H X $ in which the Higgs boson carries a significant fraction of the projectile proton momentum. The production
mechanism is based on the subprocess $(Q \bar Q) g \to H $ where the $Q \bar Q$ in the $|uud Q \bar Q>$ intrinsic heavy quark Fock state of the colliding proton has approximately
$80\%$ of the projectile protons momentum.   I discuss this in further detail in section VI.

\item{\it Eliminating the Renormalization Scale Ambiguity}\\
It is often stated that the renormalization scale of the QCD running coupling $\alpha_s(\mu^2_R) $  cannot be fixed, and thus it has to be chosen in an {\it ad hoc} fashion.  This statement is clearly false.  For example, in QED the renormalization scale is simply the photon virtuality  $q^2$ in the conventional  Gell-Mann Low scheme since this sums all vacuum polarization corrections to all orders. In fact, as in QED, the renormalization scale for perturbative QCD can be fixed unambiguously in any scheme by shifting $\mu_R$  so that all terms associated with the QCD $\beta$ function vanish.  In general, each set of skeleton diagrams has its respective scale. The result is independent of the choice of the initial renormalization scale ${\mu_R}_0$ as well as the scheme, thus satisfying the Callan-Symanzik equation and renormalization group invariance.
	
	The conventional procedure where one guesses the renormalization scale, such as $\mu^2_R = p^2_T$ and a range such as $1/2 p^2_T  < \mu^2_R <  2 p^2_T$ is clearly problematic since it depends on the choice of renormalization scheme.  This heuristic choice and range is wrong for the simplest example in QED, $e \mu \to e \mu$ scattering, since the exact answer which sums all vacuum contributions to all orders has the scale $\mu_R^2 = t$ in the Gell-Mann Low scheme [or $\mu_R^2 \simeq e^{-5/3} t$ in $\overline{MS}$  scheme ~\cite{Brodsky:1998mf}]; in fact,  the exact resummed answer at forward CM angles always lies outside the guessed heuristic range. In other cases, the choice of a heuristic scale leads to nonsensical physical results~\cite{Kramer:1990zt} or even negative cross sections at NLO ~\cite{Berger:2010zx}.
	
	Clearly the elimination of the renormalization scale ambiguity would greatly improve the precision of QCD predictions and increase the sensitivity of searches for  new physics at the LHC.
Further discussion of the renormalization scale setting problem is given in Section III.

\item{\it QCD Condensates}\\
It is conventionally assumed that the vacuum of QCD contains quark ${<0|q \bar q|0> }$ and gluon  ${<0| G^{\mu \nu} G_{\mu \nu}|0>}$ vacuum condensates, although the resulting vacuum energy density leads to a $10^{45}$  order-of-magnitude discrepancy with the
measured cosmological constant.~\cite{Brodsky:2009zd}  However, a new perspective has emerged from Bethe-Salpeter and light-front analyses where the QCD condensates are identified as ``in-hadron" condensates, rather than  vacuum entities; the Gell-Mann-Oakes-Renner  relation is still satisfied~\cite{Brodsky:2010xf}.  The ``in-hadron" condensates become realized as higher Fock states of the hadron when the theory is quantized at fixed light-front time $\tau = x^0 + x^3/c.$   I discuss this in further detail in section VII.

\item{\it Hidden Color}  \\
In nuclear physics nuclei are composites of nucleons. However, QCD provides a new perspective:~\cite{Brodsky:1976rz,Matveev:1977xt}  six quarks in the fundamental
$3_C$ representation of $SU(3)$ color can combine into five different color-singlet combinations, only one of which corresponds to a proton and
neutron.  The deuteron wavefunction is a proton-neutron bound state at large distances, but as the quark separation becomes smaller,
QCD evolution due to gluon exchange introduces four other ``hidden color" states into the deuteron
wavefunction~\cite{Brodsky:1983vf}.  The normalization of the deuteron form factor observed at large $Q^2$~\cite{Arnold:1975dd}, as well as the
presence of two mass scales in the scaling behavior of the reduced deuteron form factor~\cite{Brodsky:1976rz}, suggest sizable hidden-color
Fock state contributions  in the deuteron
wavefunction~\cite{Farrar:1991qi}. Hidden color can also play an important role in nuclear collisions involving quark distributions at high $x_F.$

\item{\it Light-Front Holography and Hadronization at the Amplitude Level}\\
A long-sought goal in hadron physics is to find a simple analytic first approximation to QCD analogous to the Schr\"odinger-Coulomb equation of atomic physics.	This problem is particularly challenging since the formalism must be relativistic, color-confining, and consistent with chiral symmetry.
de Teramond and I have shown that the correspondence between theories in a positive dilaton-modified five-dimensional anti-de Sitter space and confining field theories in physical space-time, leads to a simple
Schr\"odinger-like light-front wave equation and a remarkable one-parameter description of nonperturbative hadron dynamics~\cite{deTeramond:2005su,deTeramond:2008ht,Brodsky:2010px}. The model predicts a zero-mass pion for zero-mass quarks and a Regge spectrum of linear trajectories with the same slope in the (leading) orbital angular momentum $L$ of the hadrons and their radial  quantum number $N$.

``Light-Front Holography"~\cite{deTeramond:2008ht} allows one to map
the amplitudes $\phi(z)$ in AdS space  directly to the light-front
wavefunctions defined at fixed light-front time in 3+1 space.
The resulting Lorentz-invariant relativistic light-front wave equations are functions of  an invariant impact variable $\zeta$ which measures the separation of the quark and gluonic constituents within the hadron at equal light-front time.  This correspondence was derived by showing that the Polchinski-Strassler formula~\cite{Polchinski:2001tt}
for form factors in AdS space is equivalent to the Drell-Yan West light-front matrix element both for external electromagnetic and gravitational currents.
One then finds an exact mapping between $z$ in AdS space and the invariant impact
separation $\zeta$ in 3+1 space-time.  In the case of two-parton wavefunctions, one has  $\zeta =
\sqrt{x(1-x) b^2_\perp}$, where $b_\perp$ is the usual impact separation conjugate to
$k_\perp$ and $x = k^+/P^+$ is the light-front fraction.  This correspondence  agrees
with the intuition that $z$ is related inversely to the internal relative momentum, but the
relation $z \to \zeta$ is precise and exact. This relation also provides a direct
connection between light-front Hamiltonian equations for bound state systems and the
AdS wave equations.
One thus obtains 
a semi-classical frame-independent first approximation to the spectra and light-front wavefunctions of meson and baryon light-quark  bound states,  which in turn predicts  the
behavior of the pion and nucleon  form factors.  The theory implements chiral symmetry  in a novel way:    the effects of chiral symmetry breaking increase as one goes toward large interquark separation. The
the hadron eigenstates generally have components with different orbital angular momentum; e.g.,  the proton eigenstate in AdS/QCD with massless quarks has $L=0$ and $L=1$ light-front Fock components with equal probability.  The AdS/QCD soft-wall model also predicts the form of the non-perturbative effective coupling $\alpha_s^{AdS}(Q)$ and its $\beta$-function~\cite{Brodsky:2010ur}.

\begin{figure}[!]
 \begin{center}
\includegraphics[width=18.0cm]{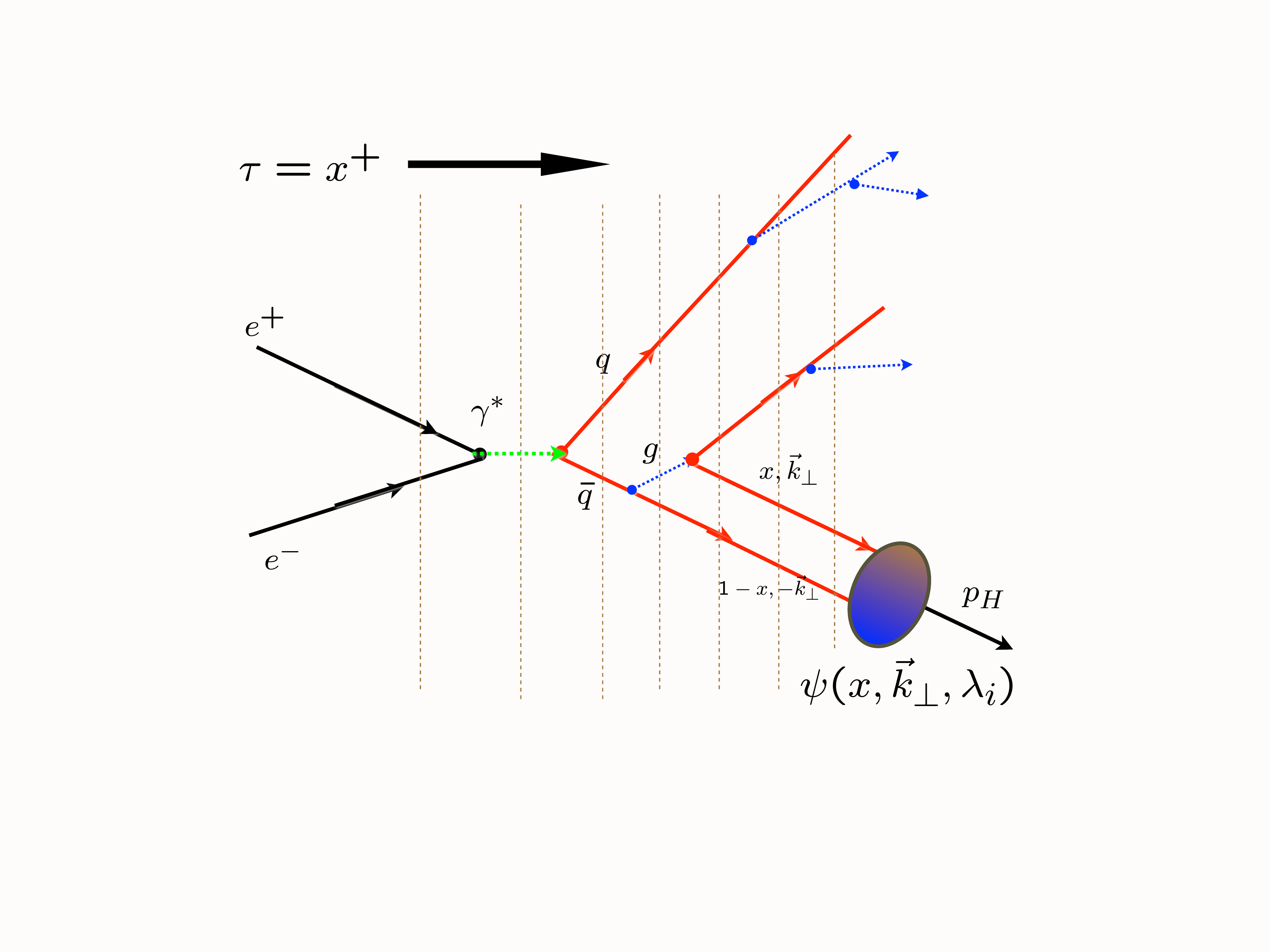}
\end{center}
  \caption{Hadronization at the amplitude level.   The off-shell $T$-matrix is computed in light-front Hamiltonian perturbation theory, evolving in light-front time $\tau = t+ z/c$.  The hadronic light-front wavefunctions convert the off-shell quarks and gluons to on-shell hadrons. }
\label{hadronize}
\end{figure}

The AdS/QCD light-front wavefunctions  obtained from AdS/QCD and Light-Front Holography provide a method for computing the hadronization of quark and gluon jets at the amplitude level~\cite{Brodsky:2008tk}.  This is illustrated for $e^+ e^- \to \gamma^* \to $ hadrons in Fig. \ref{hadronize}. An analogous method was used for QED to predict the production of relativistic antihydrogen~\cite{Munger:1993kq}.

\end{enumerate}

\section{Direct QCD Production Processes at the LHC}

It should be emphasized that the existence of dynamical higher-twist processes in which a hadron interacts directly within a hard subprocess is a prediction of
QCD.   For example, the subprocess $\gamma^* q \to \pi q,$ where the pion is produced directly through the pion's $\bar q q \to \pi$ distribution amplitude $\phi_\pi(x,Q)$ underlies  deeply virtual meson scattering $\gamma p \to \pi X.$
The corresponding timelike subprocess  $\pi q \to \gamma^*q$ dominates the Drell-Yan reaction $\pi p \to \ell^+ \ell^- X $ at high $x_F$~\cite{Berger:1979du}, thus accounting for the change in angular distribution from the canonical $1+ \cos^2\theta$ distribution, for transversely polarized virtual photons, to $\sin^2\theta$, corresponding to longitudinal photons;  the virtual photon thus becomes longitudinally polarized  at
high $x_F$, reflecting  the spin of the pion entering the direct QCD hard subprocess.
Crossing predicts
reactions where the final-state hadron appears directly in the subprocess such as $e^+ e^- \to \pi X$ at $z=1$.
The nominal power-law fall-off at  fixed $x_T$ is set by the number of elementary fields entering the hard subprocess $n_{\rm eff} = 2 n_{\rm active} -4.$   The power-law fall-off $(1-x_T)^F$ at high $x_T$ is set by the total number of spectators  $F = 2 n_{\rm spectators} -1$~\cite{Blankenbecler:1975ct}, up to spin corrections.

The direct higher-twist subprocesses, where the trigger hadron is produced  within the hard subprocess avoid the waste of same-side energy, thus allowing the target and projectile structure functions to be evaluated at the minimum values of $x_1$ and $x_2$ where they are at their maximum.
Examples of direct baryon and meson higher-twist subprocesses are: $u d \to \Lambda \bar s , u \bar d \to \pi^+ g, u g \to \pi^+ d, u \bar s \to K^+ g, u g \to K^+ s.$
These direct subprocesses involve the distribution amplitude of the hadron which has dimension  $\Lambda_{QCD}$ for mesons and
$\Lambda^2_{QCD}$ for baryons; thus these higher-twist contributions to the inclusive cross section $Ed \sigma/ d^3p$ at fixed $x_T$
nominally scale as $\Lambda^2_{QCD}/p_T^6$ for mesons and $\Lambda^4_{QCD}/p_T^8$ for baryons.

\begin{figure}[htb]
\centering
\includegraphics[width=18.0cm]{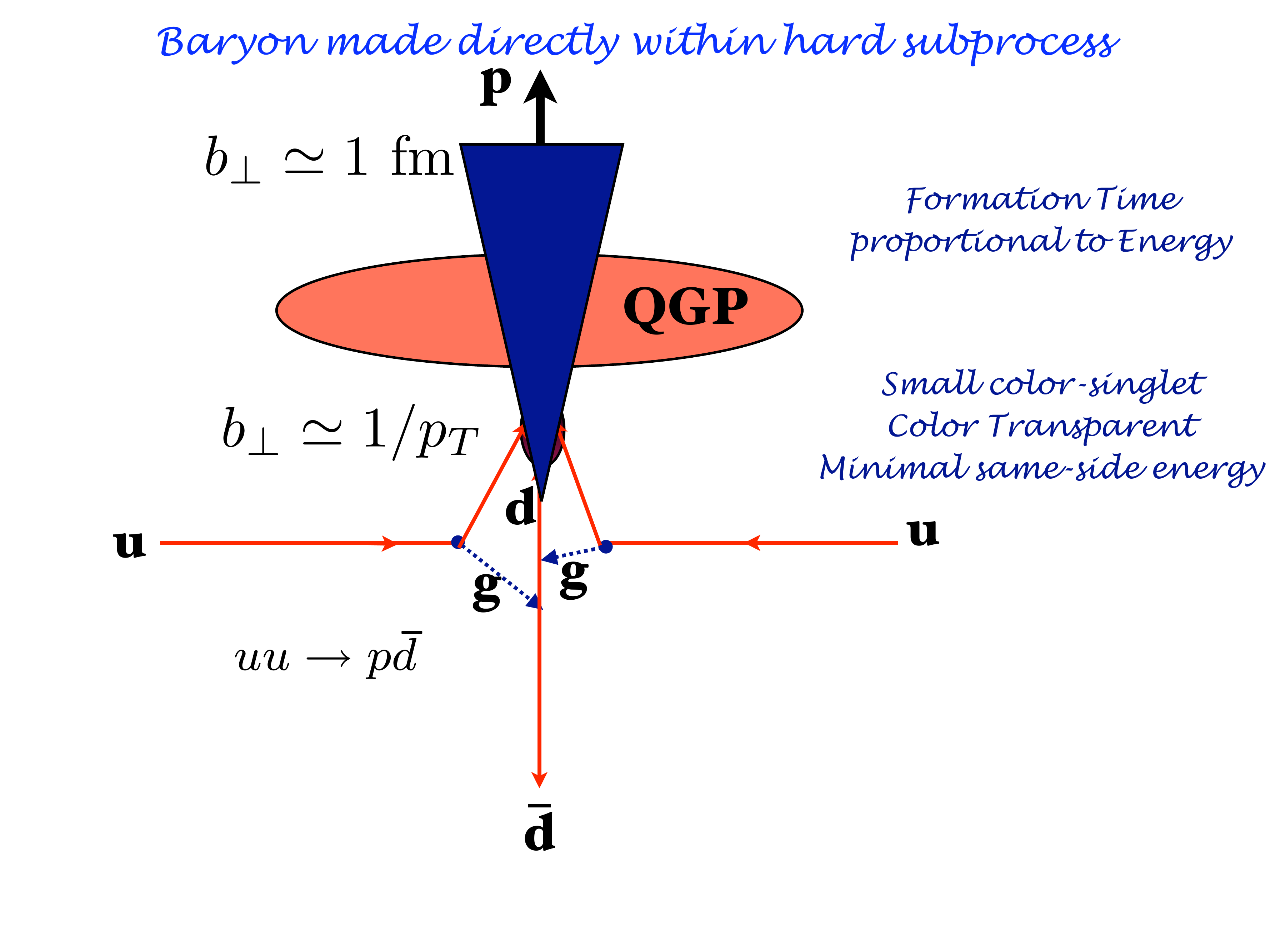}
\caption{Direct production of a proton in QCD. The proton is initially produced as a color-transparent small-size color singlet hadron. }
\label{figdirectproton}
 \end{figure}

\begin{figure}[htb]
\centering
\includegraphics[width=18.0cm]{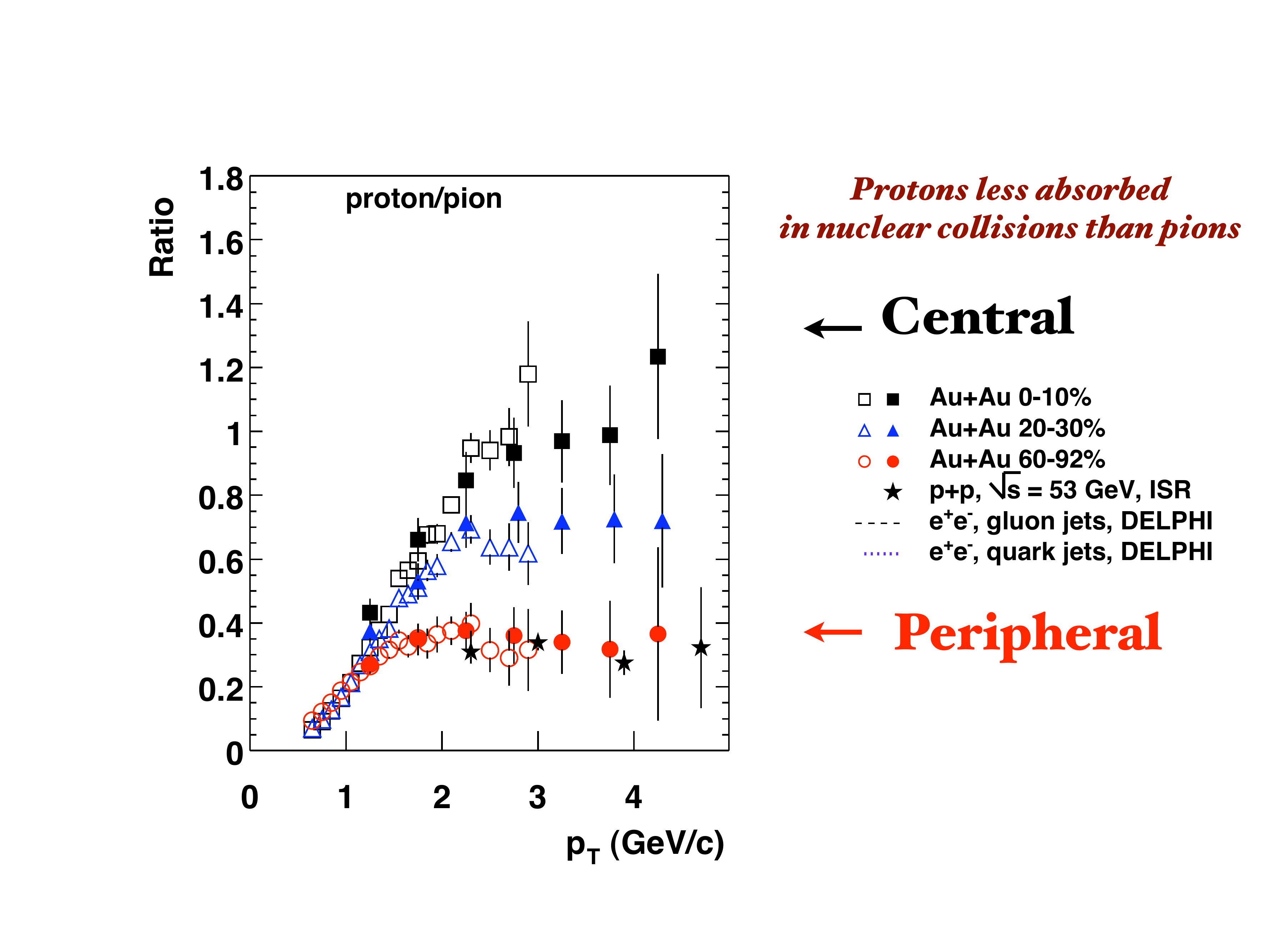}
\caption{The baryon anomaly observed by the PHENIX experiment  at RHIC~\cite{Adler:2003kg},  The anomalous rise of the proton to pion ratio with centrality at large $p_T$.  }
\label{figNew2}
 \end{figure}

\begin{figure}[htb]
\centering
\includegraphics[width=18.0cm]{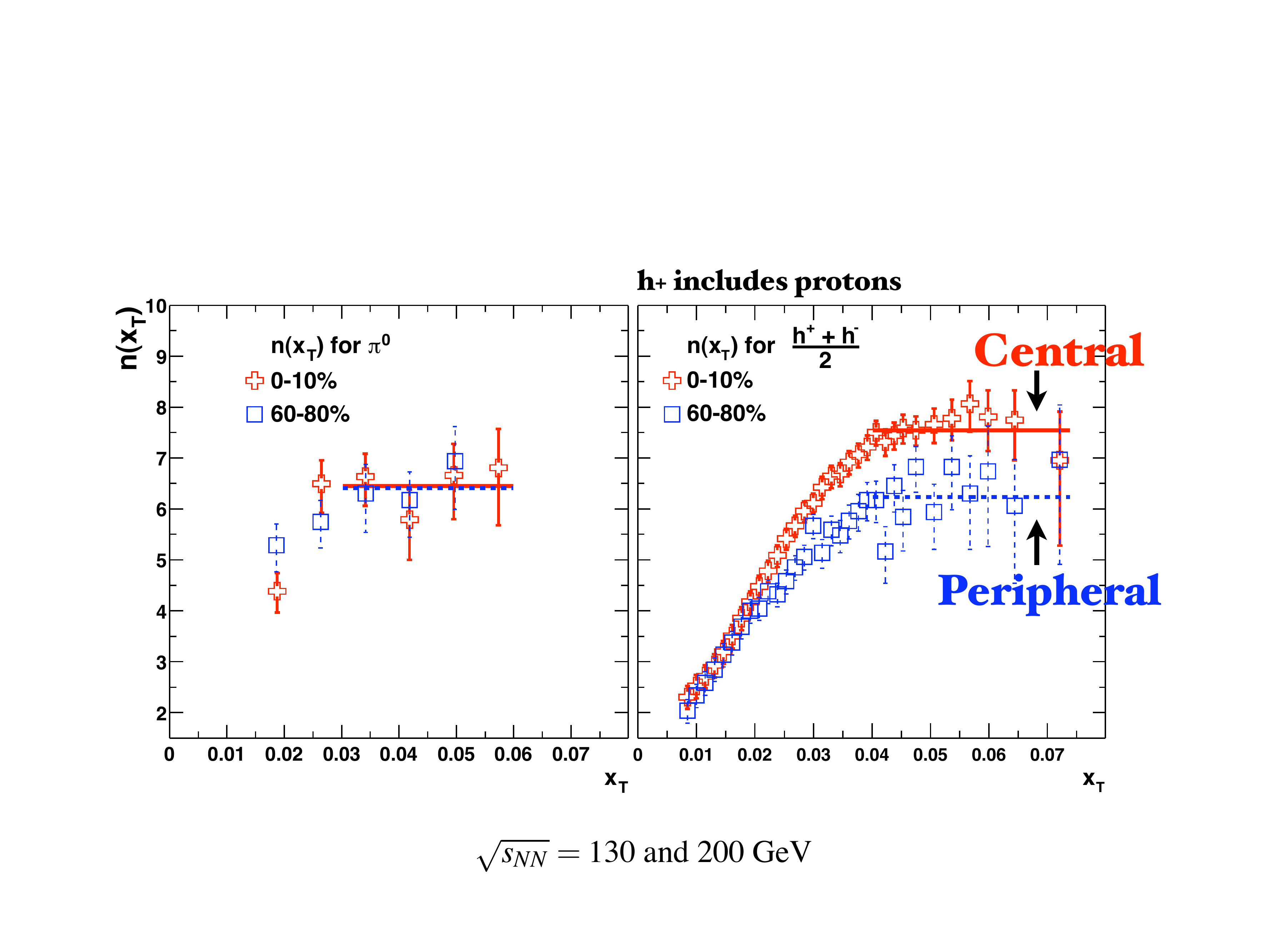}
\caption{The power-law scaling index $n_{eff}$ at fixed $x_T= {2 p_T\over \sqrt s}$  as a function of  centrality versus peripheral collisions, using spectra at $\sqrt s = 130$ GeV and
$\sqrt s =200$ GeV~\cite{Sickles:2009gy}.  The positive-charged hadron trigger is dominated by protons at high $p_T$ for central collisions, consistent with the color transparency of direct higher-twist  baryon production processes.   }
\label{figNew6}
 \end{figure}

The behavior of the single-particle inclusive cross section will be a key test of QCD at the LHC, since the leading-twist prediction for $n_{\rm eff} $ is independent of the detailed form of the structure and fragmentation functions. The fixed-$x_T$ scaling of the  proton production cross section  ${E d \sigma/ d^3p}(p p \to  p X)$ is particularly anomalous,  far from the $1/p^4_T$ to $1/p^5_T$ scaling predicted by pQCD~\cite{Brodsky:2005fz}.   See  Fig. \ref{figNew1B}.   Sickles and I~\cite{Sickles:2009gy} have argued that the anomalous features of inclusive high $p_T$ proton production is due
to hard subprocesses~\cite{Brodsky:2005fz} where the
proton is  created directly within the hard reaction, such as   $u u \to p \bar d$,
such as the mechanism illustrated in Fig. \ref{figdirectproton}.
The fragmentation of a gluon or quark jet
to a proton requires that the underlying  2 to 2 subprocess occurs at a  higher transverse momentum than the $p_T$ of the observed proton because of the fast-falling  quark-to-proton fragmentation function $D_{q \to p}(z) \sim (1-z)^3$ at high momentum fraction $z$; in contrast,  the direct subprocess is maximally energy efficient.   Such ``direct" reactions thus can explain the fast-falling power-law  falloff
observed at fixed $x_T$ and fixed-$\theta_{cm}$ at the ISR, FermiLab and RHIC~\cite{Brodsky:2005fz}.

Since the proton  is initially produced as a small-size $b_\perp \sim 1/p_T$  color-singlet state, it is ``color transparent" ~\cite{Brodsky:1988xz},   and it can thus propagate through dense nuclear matter with minimal energy loss.  In
contrast, the pions which are  produced from jet fragmentation have a  normal inelastic cross section. This provides an explanation~\cite{Brodsky:2008qp} of the RHIC
data~\cite{Adler:2003kg}, which shows a dramatic rise of the $p$ to $\pi$ ratio with increasing $p_T$ when one compares  peripheral with central heavy ion collisions, as illustrated in Fig. \ref{figNew2}.
The color transparency of the proton produced in the direct process also can explain why the index $n_{eff}$ rises with centrality, as seen in Fig. \ref{figNew6},  -- the higher-twist color-transparent subprocess dominates in the nuclear medium~\cite{Brodsky:2005fz}.   In addition,  the fact that the proton tends to be produced alone in a direct subprocess explains  why the yield of same-side hadrons along the proton trigger is diminished with increasing centrality.
Thus the QCD color transparency of directly produced baryons can  explain the ``baryon anomaly" seen in heavy-ion collisions at RHIC: the color-transparent proton state is not absorbed, but a pion produced from fragmentation is
diminished in the nuclear medium ~\cite{Sickles:2009gy}. The increase of $n_{eff}$ with centrality is consistent with the nuclear survival of direct higher-twist subprocesses for both protons and antiprotons, and to a lesser extent, for mesons.

\begin{figure}
\includegraphics[width=15.0cm]{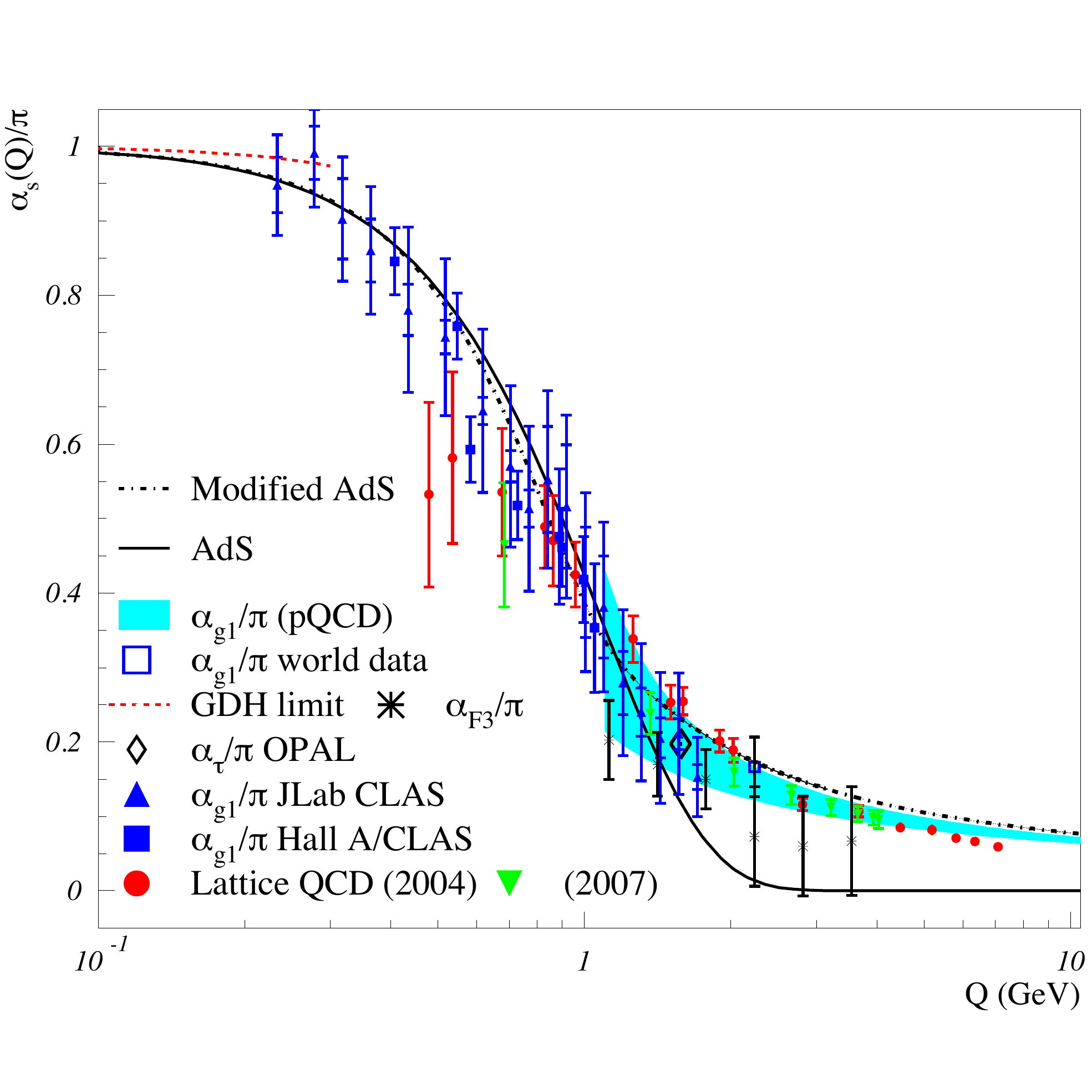}
\caption{The effective coupling  $\alpha_s(Q^2) \propto e^{-\kappa^2 Q^2}$ predicted by light-front holographic mapping  for  $\kappa = 0.54 ~ {\rm GeV}$ is compared with effective QCD couplings  extracted from
different observables and lattice results.  Details on the comparison with other effective charges are given in Ref. ~\cite{Brodsky:2010ur}.}
\label{alphas}
\end{figure}

\section{Elimination of the Renormalization Scale Ambiguity - The BLM Method}

In the BLM method~\cite{Brodsky:1982gc}, the QCD scale $\mu_R$ is chosen, just as in QED,  such that in principle all of the terms associated with the $\beta$ function are summed
into the QCD coupling~\cite{Grunberg:1991ac}
Quark loops can be used to identify the $\beta$ terms, at least to two loops
The remaining terms are identical to that of a conformal theory.  Unlike heuristic scale-setting procedures,  the BLM method gives results which are independent of the choice of renormalization scheme, as required by the transitivity property of the renormalization group.    The divergent renormalon terms of order $\alpha_s^n \beta_0^n n!$ are transferred to the physics of the running coupling.  Furthermore, one retains sensitivity to ``conformal" effects which arise in higher orders, physical effects which are not associated with QCD  renormalization.   In contrast, the factorization scale $\mu_{\rm fac}$  which sets the separation between nonperturbative dynamics of hadrons and the perturbative evolution of their parton distribution functions is arbitrary; it is unrelated to the factorization scale $\mu_R$ since it is present even in a conformal theory.

The BLM method also provides scale-fixed,
scheme-independent  connections between observables, such as the ``Generalized Crewther Relation" ~\cite{Brodsky:1995tb,Kataev:2010du}, in which Crewther's conformal prediction~\cite{Crewther:1972kn,Broadhurst:1993ru} between the Bjorken sum rule and the annihilation cross section is  effectively restored, as well as other ``Commensurate Scale Relations" ~\cite{Brodsky:1994eh,Brodsky:2000cr}. The ratio of scales in the two observables is scheme-independent and  ensures that the same number of quark flavors are active.   Such relations between observables provide high precision test of QCD. The BLM method was used in Ref. \cite{Brodsky:1998kn} to unambiguously fix the pomeron intercept predicted by the BFKL analysis. The BLM method is also correct in the Abelian limit $N_C \to 0$ at fixed $\alpha= C_F \alpha_s$~\cite{Brodsky:1997jk}, where $C_F= (N^2_C-1)/ 2 N_C $ is the Casimir constant for $SU(N_C).$

The consistent application of the BLM method to pQCD leads in most cases to multiple renormalization scales associated with different skeleton diagrams; for example in electron-electron scattering in QED  there are separate scales $\alpha(t)$ and $\alpha(u)$  for photon exchange  in the $t$ and $u$-channels respectively.  In the case of $e^+ e^- \to e^+ e^-,$ the renormalization scales are $\mu^2_R = t $ and $\mu^2_R = s$ where $\alpha(s)$ is complex.

The scale controlling the three gluon coupling in QCD is particularly interesting.  In the case of LHC processes  such as  $ p p \to Q \bar Q X,$ where the 3-gluon coupling enters at very different gluon virtualities,  the BLM scale can be shown to be proportional to $q^2_1 q^2_2 / q^2_3$ where $q^2_3$ is the largest of the gluon virtualities~\cite{Binger:2006sj,Lu:1992eq}. This scale controls the onset of the new quark flavors that enter the triangle graph.

Thus, in many cases, the actual renormalization scale is much smaller than the largest hard scale in the process. Two distinctly different scales arise in heavy quark production at threshold: the relative momentum of the quarks governing the soft gluon exchange responsible for the Coulomb potential, and a large momentum scale approximately equal to twice the quark mass for the corrections induced by transverse gluons~\cite{Brodsky:1995ds}.

Since one probes the QCD coupling at small virtuality  even at the LHC,  it is imperative to have good control on the QCD coupling in the soft domain. It is usually assumed that the QCD coupling $\alpha_s(Q^2)$ diverges at $Q^2=0$;  i.e., ``infrared slavery".  In fact, determinations from lattice gauge theory,  Bethe-Salpeter methods, effective charge measurements, gluon mass phenomena, and AdS/QCD all lead (in their respective schemes) to a finite value of the QCD coupling in the infrared~\cite{Brodsky:2010ur}.  Because of color confinement, the quark and gluon propagators vanish at long wavelength: $k < \Lambda_{QCD}$, and consequently, the quantum-loop corrections underlying the  QCD $\beta$-function --  decouple in the infrared, and  the coupling  freezes to a finite value at $Q^2 \to 0$~\cite{Brodsky:2007hb,Brodsky:2008be}.  See Fig. \ref{alphas}. This observation underlies the use of conformal methods in AdS/QCD.

Clearly the application of the BLM method to eliminate of the renormalization scale ambiguity would greatly improve the precision of QCD predictions and increase the sensitivity of searches for  new physics at the LHC.

\section{Leading-Twist Shadowing and Anti-Shadowing of Nuclear Structure Functions}

The shadowing of the nuclear structure functions: $R_A(x,Q^2) < 1 $ at small $x < 0.1 $ can be readily understood in terms of the Gribov-Glauber
theory.  Consider a two-step process in the nuclear target rest frame. The incoming $q \bar q$ dipole first interacts diffractively $\gamma^*
N_1 \to (q \bar q) N_1$ on nucleon $N_1$ leaving it intact.  This is the leading-twist diffractive deep inelastic scattering  (DDIS) process
which has been measured at HERA to constitute approximately 10\% of the DIS cross section at high energies.  The $q \bar q$ state then interacts
inelastically on a downstream nucleon $N_2:$ $(q \bar q) N_2 \to X$. The phase of the pomeron-dominated DDIS amplitude is close to imaginary,
and the Glauber cut provides another phase $i$, so that the two-step process has opposite  phase and  destructively interferes with the one-step
DIS process $\gamma* N_2 \to X$ where $N_1$ acts as an unscattered spectator. The one-step and-two-step amplitudes can coherently interfere as
long as the momentum transfer to the nucleon $N_1$ is sufficiently small that it remains in the nuclear target;  {\em i.e.}, the Ioffe
length~\cite{Ioffe:1969kf} $L_I = { 2 M \nu/ Q^2} $ is large compared to the inter-nucleon separation. In effect, the flux reaching the interior
nucleons is diminished, thus reducing the number of effective nucleons and $R_A(x,Q^2) < 1.$
The Bjorken-scaling diffractive contribution to DIS arises from the rescattering of the struck quark after it is struck  (in the
parton model frame $q^+ \le 0$), an effect induced by the Wilson line connecting the currents. Thus one cannot attribute DDIS to the physics of
the target nucleon computed in isolation~\cite{Brodsky:2002ue}.

One of the novel features of QCD involving nuclei is the {\it antishadowing} of the nuclear structure functions as observed in deep
inelastic lepton-nucleus scattering. Empirically, one finds $R_A(x,Q^2) \equiv  \left(F_{2A}(x,Q^2)/ (A/2) F_{d}(x,Q^2)\right)
> 1 $ in the domain $0.1 < x < 0.2$;  {\em i.e.}, the measured nuclear structure function (referenced to the deuteron) is larger than the
scattering on a set of $A$ independent nucleons.
Ivan Schmidt, Jian-Jun Yang, and I~\cite{Brodsky:2004qa} have extended the analysis of nuclear shadowing  to the shadowing and antishadowing of the
electroweak structure functions.  We note that there are leading-twist diffractive contributions $\gamma^* N_1 \to (q \bar q) N_1$  arising from Reggeon exchanges in the
$t$-channel~\cite{Brodsky:1989qz}.  For example, isospin--non-singlet $C=+$ Reggeons contribute to the difference of proton and neutron
structure functions, giving the characteristic Kuti-Weisskopf $F_{2p} - F_{2n} \sim x^{1-\alpha_R(0)} \sim x^{0.5}$ behavior at small $x$. The
$x$ dependence of the structure functions reflects the Regge behavior $\nu^{\alpha_R(0)} $ of the virtual Compton amplitude at fixed $Q^2$ and
$t=0.$ The phase of the diffractive amplitude is determined by analyticity and crossing to be proportional to $-1+ i$ for $\alpha_R=0.5,$ which
together with the phase from the Glauber cut, leads to {\it constructive} interference of the diffractive and nondiffractive multi-step nuclear
amplitudes.  The nuclear structure function is predicted to be enhanced precisely in the domain $0.1 < x <0.2$ where
antishadowing is empirically observed.  The strength of the Reggeon amplitudes is fixed by the Regge fit to the nucleon structure functions, so there
is little model dependence.
Since quarks of different flavors  will couple to different Reggeons; this leads to the remarkable prediction that
nuclear antishadowing is not universal; it depends on the quantum numbers of the struck quark. This picture implies substantially different
antishadowing for charged and neutral current reactions, thus affecting the extraction of the weak-mixing angle $\theta_W$.  The ratio of nuclear to nucleon structure functions is thus process dependent.   We have also identified
contributions to the nuclear multi-step reactions which arise from odderon exchange and hidden color degrees of freedom in the nuclear
wavefunction.

Schienbein et al.~\cite{Schienbein:2008ay} have recently given a comprehensive analysis of charged current deep inelastic neutrino-iron scattering, finding significant differences with the nuclear corrections for electron-iron scattering.  See Fig. \ref{figNew9}.  The nuclear effect measured in the NuTeV deep inelastic scattering charged current experiment  is distinctly different from the nuclear modification measured at SLAC and NMC in deep inelastic scattering electron and muon scattering.     This implies that part of
of the anomalous NuTeV result~\cite{Zeller:2001hh} for $\theta_W$ could be due to the non-universality of nuclear antishadowing for charged and
neutral currents.  This effect could also explain the absence of antishadowing observed in Drell-Yan reactions~\cite{Alde:1990im,Peigne:2002iw}.

A new understanding of nuclear shadowing and antishadowing has  emerged based on multi-step coherent reactions involving
leading twist diffractive reactions~\cite{Brodsky:1989qz,Brodsky:2004qa}. The nuclear shadowing of structure functions is a consequence of
the lepton-nucleus collision; it is not an intrinsic property of the nuclear wavefunction. The same analysis shows that antishadowing is {\it
not universal}, but it depends in detail on the flavor of the quark or antiquark constituent~\cite{Brodsky:2004qa}.
Detailed measurements of the nuclear dependence of individual quark structure functions are thus needed to establish the
distinctive phenomenology of shadowing and antishadowing and to make the NuTeV results definitive.  A comparison of the nuclear modification in neutrino versus anti-neutrino interactions is clearly important.  There are other ways in which this new view of antishadowing can be tested;  for example, antishadowing can also depend on the target and beam
polarization.

\begin{figure}[!]
\includegraphics[width=18cm]{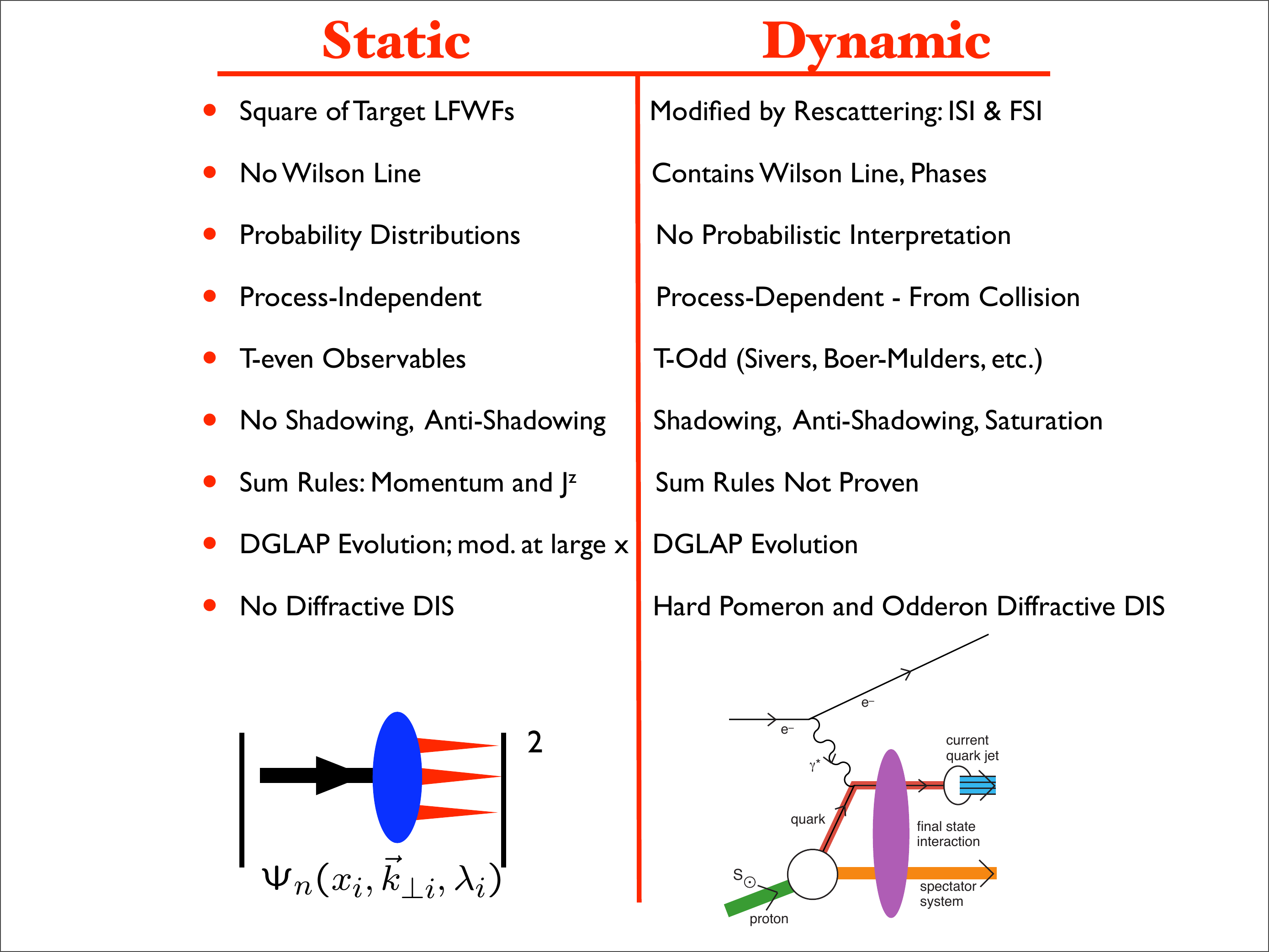}
\caption{Static versus dynamic structure functions}
\label{figNew17}
\end{figure}

\section{Dynamic versus Static Hadronic Structure Functions}
The nontrivial effects from rescattering and diffraction highlight the need for a fundamental understanding the dynamics of hadrons in QCD at the amplitude
level. This is essential for understanding phenomena such as the quantum mechanics of hadron formation, the remarkable
effects of initial and final interactions, the origins of diffractive phenomena and single-spin asymmetries, and manifestations of higher-twist
semi-exclusive hadron subprocesses. A central tool in these analyses is the light-front wavefunctions of hadrons, the frame-independent
eigensolutions of the Heisenberg equation for QCD ~  $H^{LF}|\Psi> = M^2 |\Psi>$ quantized at fixed light-front. Given the light-front
wavefunctions $\psi_{n/H}(x_i, \vec k_{\perp i}, \lambda_i )$, one can compute a large range of exclusive and inclusive hadron observables. For
example, the valence, sea-quark and gluon distributions are defined from the squares of the LFWFS summed over all Fock states $n$. Form factors,
exclusive weak transition amplitudes~\cite{Brodsky:1998hn} such as $B\to \ell \nu \pi,$ and the generalized parton
distributions~\cite{Brodsky:2000xy} measured in deeply virtual Compton scattering are (assuming the ``handbag" approximation) overlaps of the
initial and final LFWFS with $n =n^\prime$ and $n =n^\prime+2$.

It is thus important to distinguish ``static" structure functions which are computed directly from the light-front wavefunctions of  a target hadron from the nonuniversal ``dynamic" empirical structure functions which take into account rescattering of the struck quark in deep inelastic lepton scattering.
See  Fig. \ref{figNew17}.
The real wavefunctions underlying static structure functions cannot describe diffractive deep inelastic scattering nor  single-spin asymmetries, since such phenomena involve the complex phase structure of the $\gamma^* p $ amplitude.
One can augment the light-front wavefunctions with a gauge link corresponding to an external field
created by the virtual photon $q \bar q$ pair
current~\cite{Belitsky:2002sm,Collins:2004nx}, but such a gauge link is
process dependent~\cite{Collins:2002kn}, so the resulting augmented
wavefunctions are not universal.
\cite{Brodsky:2002ue,Belitsky:2002sm,Collins:2003fm}.  The physics of rescattering and nuclear shadowing is not
included in the nuclear light-front wavefunctions, and a
probabilistic interpretation of the nuclear DIS cross section is
precluded.

\section{Novel Intrinsic Heavy Quark Phenomena}

Intrinsic heavy quark distributions are a rigorous feature of QCD, arising from diagrams in which two or more gluons couple the valence quarks to the heavy quarks.
The probability for Fock states of a light hadron to have an extra heavy quark pair decreases as $1/m^2_Q$ in non-Abelian
gauge theory~\cite{Franz:2000ee,Brodsky:1984nx}.  The relevant matrix element is the cube of the QCD field strength $G^3_{\mu \nu},$  in
contrast to QED where the relevant operator is $F^4_{\mu \nu}$ and the probability of intrinsic heavy leptons in an atomic
state is suppressed as $1/m^4_\ell.$  The maximum probability occurs at $x_i = { m^i_\perp /\sum^n_{j = 1}
m^j_\perp}$ where $m_{\perp i}= \sqrt{k^2_{\perp i} + m^2_i}$; {\em i.e.}, when the constituents have minimal invariant mass and equal rapidity. Thus the heaviest constituents have the highest
momentum fractions and the highest $x_i$.  Intrinsic charm thus predicts that the charm structure function has support at large $x_{bj}$ in
excess of DGLAP extrapolations~\cite{Brodsky:1980pb}; this is in agreement with the EMC measurements~\cite{Harris:1995jx}. Intrinsic charm can
also explain the $J/\psi \to \rho \pi$ puzzle~\cite{Brodsky:1997fj}. It also affects the extraction of suppressed CKM matrix elements in $B$
decays~\cite{Brodsky:2001yt}.
The dissociation of the intrinsic charm $|uud c \bar c>$ Fock state of the proton can produce a leading heavy quarkonium state at
high $x_F = x_c + x_{\bar c}~$ in $p N \to J/\psi X $ since the $c$ and $\bar c$ can readily coalesce into the charmonium state.  Since
the constituents of a given intrinsic heavy-quark Fock state tend to have the same rapidity, coalescence of multiple partons from the projectile
Fock state into charmed hadrons and mesons is also favored.  For example, one can produce a leading $\Lambda_c$ at high $x_F$ and low $p_T$ from
the coalescence of the $u d c$ constituents of the projectile $|uud c \bar c>$  Fock state.

\begin{figure}[!]
 \begin{center}
\includegraphics[width=18.0cm]{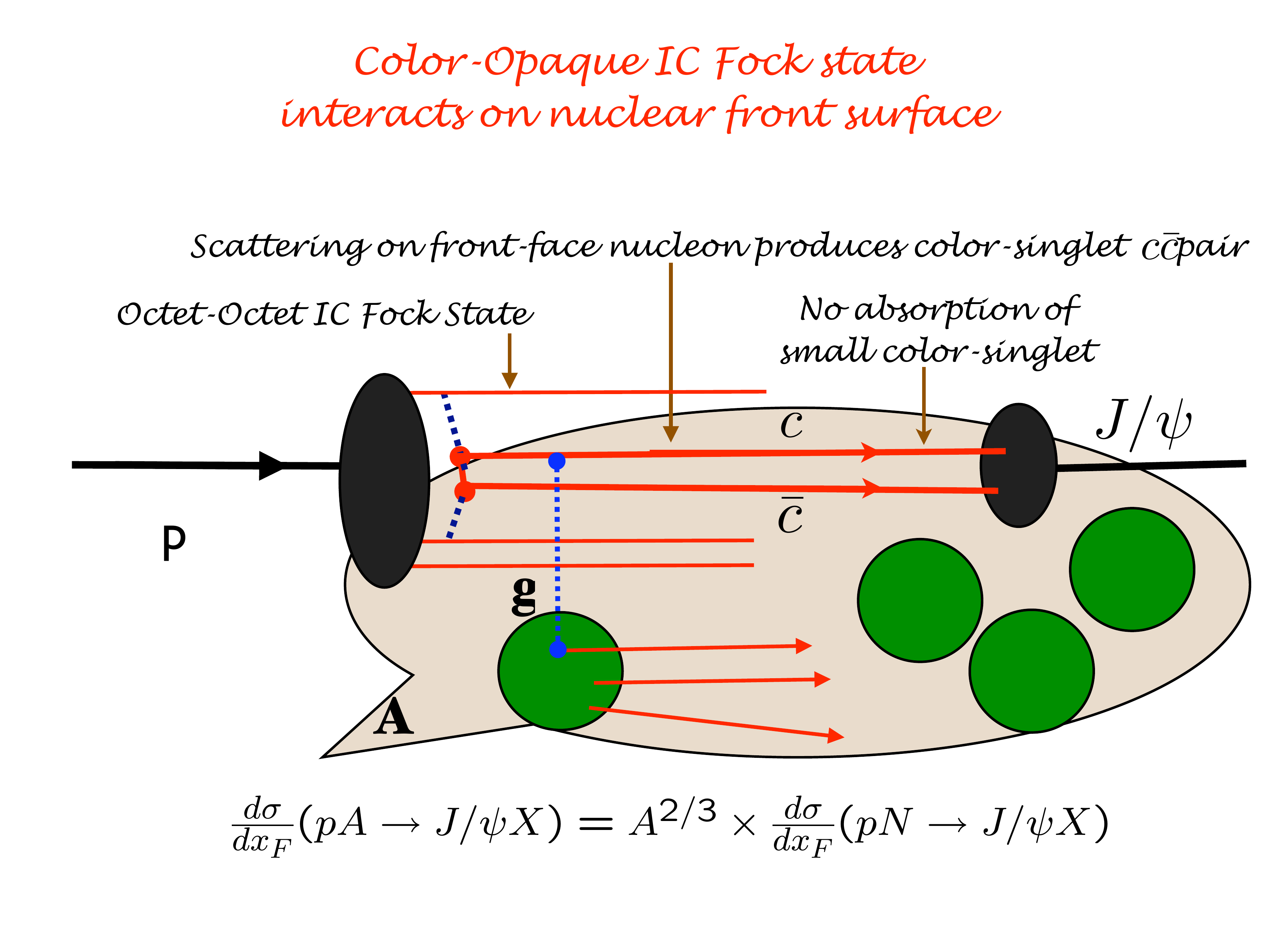}
\end{center}
  \caption{Color-Octet intrinsic charm mechanism for the nuclear dependence of $J/\psi$ production}
\label{figNew10}
\end{figure}

The operator product analysis of the IC matrix element shows that the IC Fock state has a  dominant color-octet structure:
$\vert(uud)_{8C} (c \bar c)_{8C}>$. The color octet $c \bar c$ converts to a color singlet by gluon exchange on the front surface of a nuclear target and then coalesces to a $J/\psi$ which interacts weakly through the nuclear volume~\cite{Brodsky:2006wb}.  Thus the rate for the IC component has  $A^{2/3}$ dependence corresponding to the area of the front surface.   This is illustrated in fig \ref{figNew10}.
This forward contribution is in addition to the $A^1$ contribution derived from the usual perturbative QCD
fusion contribution at small $x_F.$   Because of these two components, the cross section violates perturbative QCD factorization for hard
inclusive reactions~\cite{Hoyer:1990us}.  This is consistent with the two-component cross section for charmonium production observed by
the NA3 collaboration at CERN~\cite{Badier:1981ci} and more recent experiments~\cite{Leitch:1999ea}. The diffractive dissociation of the
intrinsic charm Fock state leads to leading charm hadron production and fast charmonium production in agreement with
measurements~\cite{Anjos:2001jr}.  The hadroproduction cross sections for  double-charm $\Xi_{cc}^+$ baryons at SELEX~\cite{Ocherashvili:2004hi} and the production of $J/\psi$ pairs at NA3 are
 consistent with the diffractive dissociation and coalescence of double IC Fock states~\cite{Vogt:1995tf}. These observations provide
compelling evidence for the diffractive dissociation of complex off-shell Fock states of the projectile and contradict the traditional view that
sea quarks and gluons are always produced perturbatively via DGLAP evolution or gluon splitting.  It is also conceivable that the observations~\cite{Bari:1991ty} of
$\Lambda_b$ at high $x_F$ at the ISR in high energy $p p$  collisions could be due to the dissociation and coalescence of the
``intrinsic bottom" $|uud b \bar b>$ Fock states of the proton.

\begin{figure}
\begin{center}
\includegraphics[width=18.0cm]{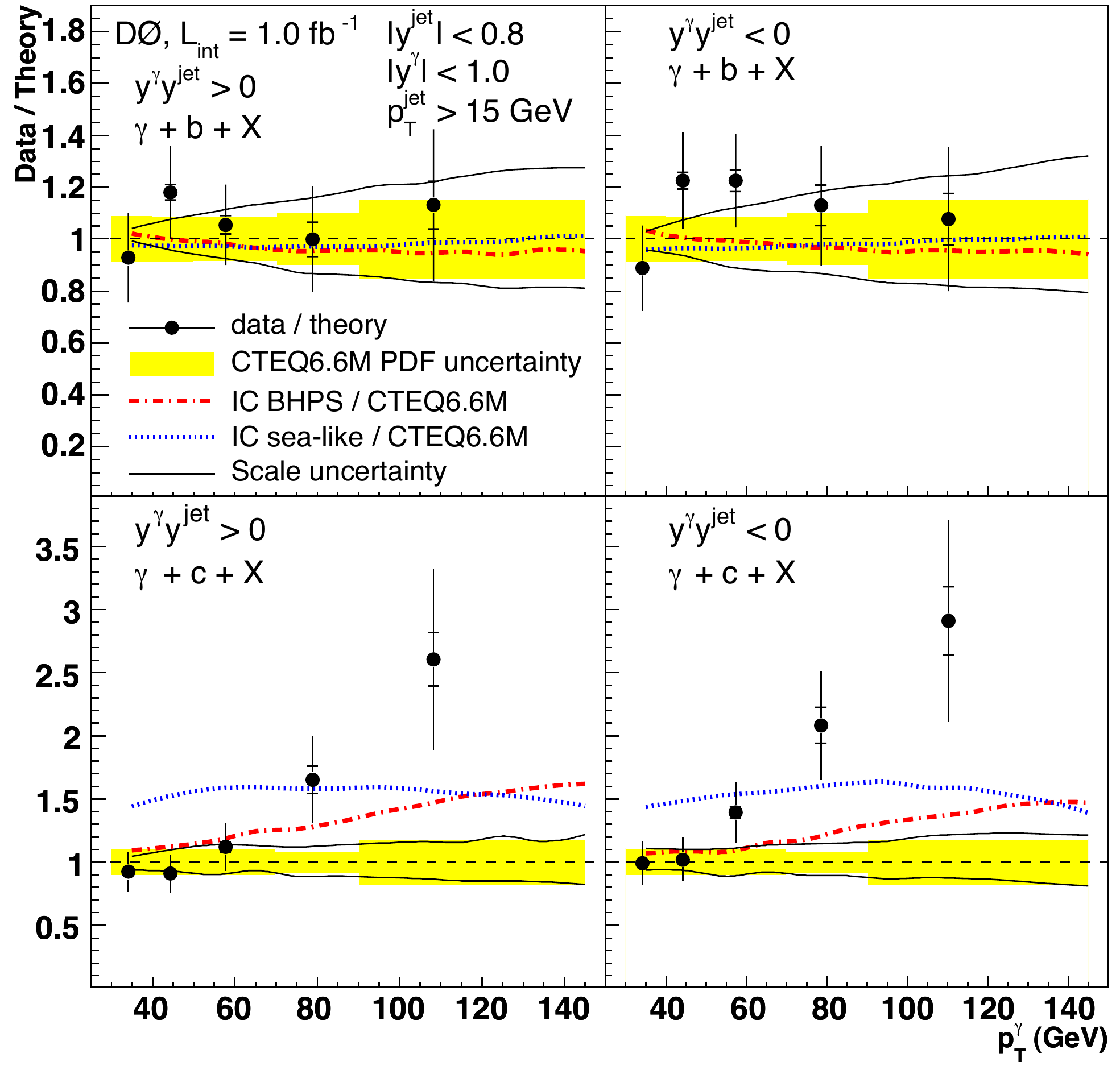}
\end{center}
\caption{The data-to-theory ratio of cross sections as a function of the photon transverse momentum
 for $ p \bar p \to \gamma  b X$ and  $ p \bar p \to \gamma  c X$  in the rapidity regions $y^{\gamma} y^\text{jet}>0$
  and $y^{\gamma} y^\text{jet}<0$.  The uncertainties for the data
  include both statistical (inner line) and full uncertainties (entire
  error bar).  Also shown are the uncertainties on the theoretical
  pQCD scales and the {\sc cteq}6.6M parameterization of the parton distribution functions (pdfs). The scale uncertainties
  are shown as dotted lines and the pdf uncertainties by the shaded
  regions.  The ratio of the standard {\sc cteq}6.6M prediction to two
  models of intrinsic charm is also shown.  From the D0 collaboration~\cite{Abazov:2009de}.  }
\label{figratio}
\end{figure}

As emphasized by Lai, Tung, and Pumplin~\cite{Pumplin:2007wg}, there are strong indications that the structure functions used to model charm
and bottom quarks in the proton at large $x_{bj}$ have been underestimated, since they ignore intrinsic heavy quark fluctuations of
hadron wavefunctions.  The anomalous growth of the $p \bar p \to \gamma  c X$ inclusive cross section observed by D0 collaboration~\cite{Abazov:2009de} at the Tevatron and shown in Fig. \ref{figratio} also suggests that the charm  distribution has been underestimated at $x> 0.1,$ beyond
what is included in the {\sc cteq}6.6M  pdfs.

The neglect of the intrinsic-heavy quark component  in the proton structure function will lead to an incorrect assessment of the gluon distribution at large $x$ if it is assumed that sea quarks always arise from gluon splitting.

The intrinsic charm and bottom distributions at high $x$ in the proton structure functions also lead to novel heavy hadron production processes such as anomalous production of charm and bottom jets at high $p_T$ at the LHC, as well as novel mechanisms for Higgs and $Z^0$ production at high $x_F$
~\cite{Brodsky:2007yz,Brodsky:2006wb}

 It is astonishing that the original EMC experiment which first observed a large signal for charm at large $x$ in 1983  has never been repeated.  It is thus critical for experiments such as COMPASS to definitively establish the phenomenology of the charm structure function at
large $x_{bj}.$

\section {Vacuum Effects and Light-Front Quantization}

The vacuum in quantum field theories is remarkably simple in light-front quantization because of the restriction $k^+ \ge 0.$   For example in QED,  vacuum graphs such as $e^+ e^- \gamma $  which are normally associated with a zero-point energy do not arise in the light-front vacuum. In the Higgs theory, the usual Higgs vacuum expectation value is replaced with a $k^+=0$ zero mode;~\cite{Srivastava:2002mw} however, the resulting phenomenology is identical to the standard analysis.

The usual assumption that non-zero vacuum condensates exist and possess a measurable reality has long been recognized as posing a conundrum for the light-front formulation of QCD.   In the light-front formulation of QCD, the ground-state is a structureless Fock space vacuum, in which case it would seem to follow that dynamical chiral symmetry breaking (CSB)  is impossible.  In fact,  as first argued by Casher and Susskind \cite{Casher:1974xd}, dynamical CSB must be a property of hadron wavefunctions, not of the vacuum. (They used an infinite momentum framework which is equivalent to the front-form.) This thesis has also been explored in a series of recent articles
\cite{Brodsky:2008tk, Brodsky:2008be, Brodsky:2009zd}.

It is widely held that quark and gluon vacuum condensates have a physical existence, independent of  hadrons, as measurable spacetime-independent configurations of QCD's elementary degrees-of-freedom in a hadron-less ground state. However, a  non-zero spacetime-independent QCD vacuum condensate poses a critical dilemma for gravitational interactions, because it would lead to a cosmological constant some 45 orders of magnitude larger than observation.  As noted  in Ref. \cite{Brodsky:2009zd}, this conflict is avoided if the strong interaction condensates are properties of the light-front wavefunctions of the hadrons, rather than the hadron-less ground state of QCD.

Conventionally, the quark and gluon condensates are considered to be properties
of the QCD vacuum and hence to be constant throughout spacetime.
A new perspective on the nature of QCD
condensates $\langle \bar q q \rangle$ and $\langle
G_{\mu\nu}G^{\mu\nu}\rangle$, particularly where they have spatial and temporal
support,
has recently been presented.~\cite{Brodsky:2010xf,Brodsky:2008be,Brodsky:2009zd,Brodsky:2008xm,Brodsky:2008xu}
The spatial support of condensates is restricted to the interior
of hadrons, since condensates in QCD arise due to the interactions of quarks and
gluons which are confined within hadrons.
For example, consider a meson consisting of a light quark $q$ bound to a heavy
antiquark, such as a $B$ meson.  One can analyze the propagation of the light
$q$ in the background field of the heavy $\bar b$ quark.  Solving the
Dyson-Schwinger equation for the light quark one obtains a nonzero dynamical
mass and evidently a nonzero value of the
condensate $\langle \bar q q \rangle$.  But this is not a true vacuum
expectation value; in fact, it is the matrix element of the operator $\bar q q$
in the background field of the $\bar b$ quark.  

The change in the (dynamical)
mass of the light quark in a bound state is somewhat reminiscent of the
energy shift of an electron in the Lamb shift, in that both are consequences of
the fermion being in a bound state rather than propagating freely.
It is clearly important to use the equations of motion for confined quarks
and gluon fields when analyzing current correlators in QCD, not free
propagators, as has often been done in traditional analyses. Since the distance between the
quark and antiquark cannot become arbitrarily large, one cannot create a quark
condensate which has uniform extent throughout the universe.
Thus in a fully self-consistent treatment of a QCD bound state, the condensate phenomenon occurs in the background field of the $\bar b$-quark, whose influence on light-quark propagation is primarily concentrated in the far infrared and whose presence ensures the manifestations of light-quark dressing are gauge invariant.

In the case of the pion, one can show from the Bethe-Salpeter equation that the vacuum quark condensate  that appears in the  Gell-Mann-Oakes-Renner formula, is, in fact, a chiral-limit value of an `in-pion' condensate~\cite{Maris:1997tm,Brodsky:2010xf}. This condensate is no more a property of the ``vacuum'' than the pion's chiral-limit leptonic decay constant.

One can connect the Bethe-Salpeter formalism to the light-front formalism, by fixing the light-front time $\tau$. This then leads to the Fock state expansion.  In fact,  dynamical CSB in the light-front formulation, expressed via `in-hadron' condensates, can be shown to be connected with sea-quarks derived from higher Fock states.  This solution is similar  to that discussed in Ref. \cite{Casher:1974xd}.
Moreover, Ref. \cite{Langfeld:2003ye} establishes the equivalence of all three definitions of the vacuum quark condensate: a constant in the operator product expansion,~\cite{Lane:1974he,Politzer:1976tv} via the Banks-Casher formula,~\cite{Banks:1979yr} and the trace of the chiral-limit dressed-quark propagator.

\section{Conclusions}
I have reviewed a number of QCD topics where conventional wisdom relevant to hadronic physics at the LHC  has been challenged.

For example, the initial-state and final-state interactions of the  quarks and gluons entering perturbative QCD hard-scattering subprocesses lead to the breakdown of traditional concepts of factorization and universality at leading twist.    These soft-gluon rescattering, which are associated with the Wilson line of the propagating partons, lead to Bjorken-scaling single-spin asymmetries, diffractive deep inelastic scattering, the breakdown of
the Lam-Tung leading twist relation in Drell-Yan reactions, as well as nuclear shadowing.
Furthermore, the Gribov-Glauber theory applied to the antishadowing domain predicts that nuclear structure functions are not  universal, but instead depend on the flavor quantum numbers of each quark and antiquark, thus explaining the anomalous nuclear dependence recently observed in deep-inelastic neutrino scattering.

Surprisingly,
isolated hadrons can be produced at large transverse momentum at a significant rate at the LHC directly within a hard higher-twist QCD subprocess, rather than from jet fragmentation.   Such ``direct"  processes  can explain the observed deviations from perturbative QCD predictions in measurements of  inclusive hadron cross sections at  fixed $x_T= 2p_T/\sqrt s$,  as well as the ``baryon anomaly", the  anomalously large proton-to-pion ratio seen in
high centrality heavy-ion collisions.

The intrinsic charm contribution to the proton structure function  at high $x$  can explain the anomalously large rate for high $p_T$ photon plus charm jet events observed by D0 at the Tevatron. Intrinsic charm and bottom distributions also imply anomalously large production of charm and bottom jets at high $p_T$ at the LHC,  as well as a novel mechanism for Higgs and $Z^0$ production at high $x_F.$

The correspondence  between theories in a warped anti-de Sitter space and light-front quantization in physical space-time is very powerful, and
leads to much insight into QCD dynamics, including a nonperturbative QCD running
coupling and a remarkably accurate relativistic LF Schrodinger equation which
reproduces much of light-quark spectroscopy and dynamics,  using a soft-wall model of
with a positive sign dilaton.

Other novel features of QCD have also been discussed, including the consequences of confinement for quark and gluon condensates and the implications for the QCD contribution to the cosmological constant.

I have also emphasized that setting the renormalization scale of the QCD coupling using
the scheme-independent BLM method will greatly improve the precision of QCD predictions, and thus greatly increase the sensitivity of searches for new physics at the LHC.

$$ $$

\vspace{10pt}

\noindent{\bf Acknowledgments}

\vspace{5pt}

Invited talk, presented at the 5th Workshop on High $P_T$ Physics  at the LHC held at the Instituto de Ciencias Nucleares of the  Universidad National Automata de Mexico in Mexico City, September 27-October 1, 2010.  I am grateful to the organizers for their invitation to this meeting, and I thank all of my collaborators whose work has been cited in this report.
This research was supported by the Department
of Energy,  contract DE--AC02--76SF00515.
SLAC-PUB-14302.

\end{document}